\documentclass[aos,MSNbibl,nameyear,dvips]{arximspdf}
\usepackage{graphicx}
\doi{10.1214/12-AOS1078} 
\volume{41}
\issue{3}
\pubyear{2013}
\firstpage{1085}
\lastpage{1110}

\makeatletter

\newcommand{\rright}{\right}
\newcommand{\lleft}{\left}

\newtheorem{theorem}{Theorem}[section]
\newtheorem{lemma}[theorem]{Lemma}
\newtheorem{corollary}[theorem]{Corollary}

\makeatother

\begin{document}
\begin{frontmatter}

\title{Maximum lilkelihood estimation in the $\beta$-model\thanksref{T1}}
\runtitle{Maximum lilkelihood estimation in the $\beta$-model}

\thankstext{T1}{Supported in part by Grant
FA9550-12-1-0392 from the U.S. Air Force Office of Scientific Research
(AFOSR) and the Defense Advanced Research Projects Agency (DARPA), NSF
Grant DMS-06-31589, and by a grant from the Singapore National Research
Foundation (NRF) under the Interactive \& Digital Media Programme
Office to the Living Analytics Research Centre (LARC).}

\begin{aug}
\author[A]{\fnms{Alessandro} \snm{Rinaldo}\corref{}\ead[label=e1]{arinaldo@cmu.edu}},
\author[B]{\fnms{Sonja}~\snm{Petrovi\'c}\ead[label=e2]{petrovic@stat.psu.edu}}
\and\break
\author[C]{\fnms{Stephen~E.}~\snm{Fienberg}\ead[label=e3]{fienberg@stat.cmu.edu}}
\runauthor{A. Rinaldo, S. Petrovi\'c and S. E. Fienberg}
\affiliation{Carnegie Mellon University, Pennsylvania State University
and Carnegie~Mellon~University}
\address[A]{A. Rinaldo\\
Department of Statistics\\
Carnegie Mellon University\\
5000 Forbes Avenue\\
Pittsburgh, Pennsylvania 15213\\
USA\\
\printead{e1}}
\address[B]{S. Petrovi\'c\\
Department of Statistics\\
Pennsylvania State University\\
326 Thomas Building\\
University Park, Pennsylvania 16802\\
USA\\
\printead{e2}}
\address[C]{S. E. Fienberg\\
Department of Statistics\\
Machine Learning Department\\
Cylab\\
Heinz School\\
Carnegie Mellon University\\
5000 Forbes Avenue\\
Pittsburgh, Pennsylvania 15213\\
USA\\
\printead{e3}} 
\end{aug}

\received{\smonth{9} \syear{2011}}
\revised{\smonth{12} \syear{2012}}

%
\begin{abstract}
We study maximum likelihood estimation for the statistical model for
undirected random graphs, known as the $\beta$-model, in which the
degree sequences are minimal sufficient statistics. We derive
necessary and sufficient conditions, based on the polytope of degree
sequences, for the existence of the maximum likelihood estimator (MLE)
of the model parameters. We characterize in a combinatorial fashion
sample points leading to a nonexistent MLE, and nonestimability of the
probability parameters under a nonexistent MLE. We formulate conditions
that guarantee that the MLE exists with probability tending to one as
the number of nodes increases.
\end{abstract}

%
\begin{keyword}[class=AMS]
\kwd{62F99}
\end{keyword}
\begin{keyword}
\kwd{$\beta$-model}
\kwd{polytope of degree sequences}
\kwd{random graphs}
\kwd{maximum likelihood estimator}
\end{keyword}

\end{frontmatter}

\section{Introduction}

Many statistical models for the representation and analysis of network
data rely on information contained in the \textit{degree sequence}, the
vector of node degrees of the observed graph. Node degrees
not only quantify the overall connectivity of the network, but also
reveal other potentially more refined features of interest. The study
of the degree sequences and, in particular, of the degree
distributions of real networks is a classic topic in network analysis,
which has received extensive treatment in the statistical literature
[see, e.g., \citet{HL81,fienwass1981,fienmeyewass1985}], the
physics literature [see, e.g.,
\citet{NSW01,AB02,N03,PN04,Newman-Barabasi-Watts06,FOSTER07,Will09}]
as well as in the social network literature [see, e.g., \citet
{Robins08,Goodreau07,HM07}
and references therein]. See also the
monograph by \citet{gzfa10} and the books by \citet{Kolaczyk09},
\citet{Cohen-Hvalin10} and \citet{Newman10}.

The simplest instance of a statistical network model based exclusively
on the node degrees is the exponential family of probability
distributions for undirected random graphs with the degree sequence as
its natural sufficient statistic. This is in fact a simpler, undirected
version of the broader class of statistical models for directed
networks known as the $p_1$-models, introduced by \citet{HL81}. We
will refer to this model as the \textit{beta} model (henceforth the
$\beta
$-model), a name recently coined by \citet{CDS10}, and refer to
\citet
{BD09} for details and extensive references.

Despite its apparent simplicity and popularity, the $\beta$-model, much
like most network models, exhibits nonstandard statistical features,
since its complexity, measured by the dimension of the parameter
space, increases with the size of the graph. Lauritzen (\citeyear{LAU03,LAU08})
characterized $\beta$-models as the natural models for representing
exchangeable binary arrays that are \textit{weakly summarized}, that is,
random arrays whose distribution only depends on the row and column
totals. More recently, \citet{CDS10} conducted an analysis of the
asymptotic properties of the $\beta$-model, including existence and
consistency of the maximum likelihood estimator (MLE) as the dimension
of the network increases, and provided a simple algorithm for
estimating the natural parameters. They also characterized the graph
limits, or \textit{graphons} [see \citet{LS06}], corresponding to a
sequence of $\beta$-models with given degree sequences [for a
connection between the theory of graphons and exchangeable arrays
see \citet{DJ07}]. Concurrently, \citet{BH10} explored the asymptotic
behavior of sequences of random graphs with given degree sequences, and
studied a different mode of stochastic convergence. Among other things,
they show that, as the size of the network increases and under a
``tameness'' condition, the number of edges of a uniform graph with
given degree sequence converges in probability to the number of edges
of a random graph drawn from a $\beta$-model parametrized by the MLE
corresponding to degree sequence. \citet{Y1} and \citet{Y2} derived
asymptotic conditions for uniform consistency and asymptotic normality
of the MLE of the $\beta$-model, and asymptotic normality of the
likelihood ratio test for homogeneity of the model parameters.
\citet{PW12} consider a general class of models for network data
parametrized by node-specific parameters, of which the $\beta$-model is
a special case. The authors derive nonasymptotic conditions under
which the MLEs of model parameters exist and can be well approximated
by simple estimators.

In an attempt to avoid the reliance on asymptotic methods, whose
applicability to network models remains largely unclear
[see, e.g., \citet{HAB81}], several researchers have turned to
exact inference
for the $\beta$-model, which hinges upon the nontrivial task of
sampling from the set of graphs with a given degree sequence.
\citet{BD09} developed and analyzed a sequential importance sampling
algorithm for generating a random graph with the prescribed degree
sequence [see also \citet{VL05} for a different algorithm].
\citet{HT10} and \citet{markovjapan} tackled the same task
using more
abstract algebraic methods, and \citet{p1markov} studied Markov
bases for the more general $p_1$ model.

In this article we study the existence of the MLE for the parameters of
the $\beta$-model under a more general sampling scheme in which each
edge is observed a fixed number of times (instead of just once, as in
previous works) and for increasing network sizes. We view the issue of
existence of the MLE as a natural measure of the intrinsic statistical
difficulty of the $\beta$-model for two reasons.
First, existence of the MLE is a natural minimum requirement for
feasibility of statistical inference in discrete exponential families,
such as the $\beta$-model: nonexistence of the MLE is in fact
equivalent to nonestimability of the model parameters, as illustrated
in \citet{MLE}. Thus, establishing conditions for existence of the MLE
amounts to specifying the conditions under which statistical inference
for these models is fully possible. Second, under the asymptotic
scenario of growing network sizes, existence of the MLE will provide a
natural measure of sample complexity of the $\beta$-model and will
indicate the asymptotic scaling of the model parameters for which
statistical inference is viable.

Though \citet{CDS10} and \citet{BH10}\setcounter{footnote}{1}\footnote{In the
analysis of
\citet{BH10}, the \textit{maximum entropy matrix} associated to a degree
sequence is in fact exactly the MLE corresponding to the observed
degree sequence. This is a well-known property of linear exponential
families; see, for example, \citet{CT91}, Chapter 11.} also
considered the existence of the MLE, our analysis differs substantially
from theirs in that it is rooted in the statistical theory of discrete
linear exponential families and relies in a fundamental way on the
geometric properties of these families
[see, in particular, \citet{ERGM,GEYER09}]. Our contributions are
as follows:

\begin{itemize}
\item We provide explicit necessary and sufficient conditions for
existence of the MLE for the $\beta$-model that are based on the
polytope of degree sequences, a~well-studied polytope arising in the
study of threshold graphs; see \citet{MP95}. In contrast, the
conditions of \citet{CDS10} are only sufficient. We then show that
nonexistence of the MLE is brought on by certain forbidden patterns of
extremal network configurations, which we characterize in a
combinatorial way. Furthermore, when the MLE does not exist, we can
identify exactly which probability parameters are estimable.
\item We
use the properties of the polytope of degree sequences to formulate
geometric conditions that allow us to derive finite sample bounds on
the probability that the MLE does not exist. Our asymptotic results
improve analogous results of \citet{CDS10} and our proof is both
simpler and more direct. Furthermore, we show that the tameness
condition of \citet{BH10} is stronger than our conditions for existence
of the MLE.
\item Our analysis is not specific to the $\beta$-model
but, in fact, follows a principled way for detecting nonexistence of
the MLE and identifying nonestimable parameters that is based on
polyhedral geometry and applies more generally to discrete models. We
illustrate this point by analyzing other network models that are
variations or generalizations of the $\beta$-model: the $\beta$-model
with random numbers of edges, the Rasch model, the Bradley--Terry model
and the $p_1$ model. Due to space limitations, the details of these
additional analyses are contained in the supplementary material
[\citet{RinPetFie}].
\end{itemize}

While this is a self-contained article, the results derived here are
best understood as applications of the geometric and combinatorial
properties of log-linear models under product-multinomial sampling
schemes, as detailed in \citet{MLE} and its supplementary
material, to
which we refer the reader for further details as well as for practical
algorithms.

The article is organized in the following way. Section~\ref{secintro}
introduces the $\beta$-model and establishes the exponential family
parametrization that is key to our analysis. In Section
\ref{secexistence} we derive necessary and sufficient conditions for
existence of the MLE of the $\beta$-model parameters and characterize
parameter estimability under a nonexistent MLE. These results are
further discussed with examples in Section~\ref{secapplic}. In
Section~\ref{secasymptotics} we provide sufficient conditions on the
expected degree sequence guaranteeing that, with high probability as
the network size increases, the MLE exists. Finally, in Section \ref
{secdiscussion} we indicate possible extensions of our work and
briefly discuss some of the computational issues directly related to
detecting nonexistence of the MLE and parameter estimability.

We will assume throughout some familiarity with basic concepts from
polyhedral geometry [see, e.g., \citet{SCH98}] and the theory of
exponential families; see, for example, \citet{BN78,BROWN86}.

\section{\texorpdfstring{The (generalized) $\beta$-model}
{The (generalized) beta-model}}\label{secintro}

In this section we describe the exponential family parametrization of a
simple generalization of the $\beta$-model, which, with slight abuse
of notation, we will refer to as the $\beta$-model as well.

We are concerned with modeling the occurrence of edges in a simple
undirected random graph with node set $\{1,\ldots,n\}$.
The statistical experiment consists of recording, for each pair of
nodes $(i,j)$ with $i < j$, the number of edges appearing in $N_{i,j}$
i.i.d. samples, where the integers $\{ N_{i,j}, i<j \}$ are
deterministic and positive (we can relax both the nonrandomness and
positivity assumptions).
Thus, in our setting we allow for the possibility that each edge in the
network be sampled a different number of times, a realistic feature
that makes the model more flexible.
For $i < j$, we denote by $x_{i,j}$, the number of times we observe the
edge $(i,j)$ and, accordingly, by $x_{j,i}$ the number of times edge
$(i,j)$ is missing. Thus, for all $(i,j)$,
\[
x_{i,j} + x_{j,i} = N_{i,j}.
\]
We model the observed edge counts $\{ x_{i,j}, i < j\}$ as draws from
mutually independent binomial distributions, with $x_{i,j} \sim
\operatorname
{Bin}(N_{i,j},p_{i,j})$, where $p_{i,j} \in(0,1)$ for each $i < j$.

Data arising from such an experiment has a representation in the form
of a $n \times n$ contingency table with empty diagonal cells and whose
$(i,j)$th cell contains the count $x_{i,j}$, $i \neq j$.
For modeling purposes, however, we need only consider the
upper-triangular part of this table. Indeed, since, given $x_{i,j}$,
the value of $x_{j,i}$ is determined by $N_{i,j} - x_{i,j}$, we can
represent the sample space more parsimoniously as the following subset
of $\mathbb{N}^{n \choose2}$:
\[
\mathcal{S}_n:= \bigl\{ x_{i,j} \dvtx i < j \mbox{ and }
x_{i,j} \in\{0,1,\ldots,N_{i,j} \} \bigr\}.
\]
We index the coordinates $\{ (i,j) \dvtx i < j \}$ of any point in
$\mathcal{S}_n$ lexicographically.

In the $\beta$-model, we parametrize the ${n \choose2}$ edge
probabilities by points $\beta\in\mathbb{R}^n$ as follows. For each
$\beta\in\mathbb{R}^n$, the probability parameters are uniquely
determined as
%
%
\begin{equation}
\label{eqpij} p_{i,j} = \frac{e^{\beta_i+ \beta_j}}{1 + e^{\beta_i +
\beta_j}} \quad\mbox{and}\quad p_{j,i} =
1 - p_{i,j} = \frac{1}{1 + e^{\beta
_i + \beta_j}}\qquad\forall i \neq j
\end{equation}
or, equivalently, in terms of log-odds,
%
%
\begin{equation}
\label{eqodds} \log\frac{p_{i,j}}{1 - p_{i,j}} = \beta_i +
\beta_j\qquad\forall i \neq j.
\end{equation}
The magnitude and sign of $\beta_i$ quantifies the propensity of node
$i$ to have ties: the degree of node $i$ is expected to be large
(small) if $\beta_i$ is positive (negative) and of large magnitude.
Thus the $\beta$-model is the natural heterogenous version of the
well-known Erd{\H{o}}s--R\'enyi random graph model [\citet{ErdoReny1959}].
For a discussion of this model and its generalizations see
\citet{gzfa10}.

For a given choice of $\beta$, the probability of observing the vector
of edge counts $x \in\mathcal{S}_n$ is
%
%
\begin{equation}
\label{eqlik} p_\beta(x) = \prod_{i < j}
\pmatrix{N_{i,j}
\cr
x_{i,j}} p_{i,j}^{x_{i,j}}(1
- p_{i,j})^{N_{i,j}- x_{i,j}}
\end{equation}
with the probability values $p_{i,j}$ satisfying (\ref{eqpij}).
Simple algebra allows us to rewrite this expression in exponential
family form as
%
%
\begin{equation}
\label{eqdens} p_\beta(x) = \exp\Biggl\{ \sum
_{i=1}^n d_i \beta_i -
\psi(\beta) \Biggr\} \prod_{i < j} \pmatrix{N_{i,j}
\cr
x_{i,j}},
\end{equation}
where the coordinates of the vector of the minimal sufficient
statistics $d = d(x) \in\mathbb{N}^n$ are
%
%
\begin{equation}
\label{eqsuffstat} d_i = \sum_{j<i}x_{j,i}
+ \sum_{j>i} x_{i,j},\qquad i=1,\ldots,n,
\end{equation}
and the log-partition function is $\psi(\beta) = \sum_{i < j}
N_{i,j} \log( 1 + e^{\beta_i + \beta_j} )$.

Note that $e^{\psi(\beta)} < \infty$ for all $\beta\in\mathbb{R}^n$,
so $\mathbb{R}^n$ is the natural parameter space of the full and steep
exponential family with support $\mathcal{S}_n$
[see, e.g., \citet{BN78}] and densities given by the exponential
term in
(\ref{eqdens}).

\subsection*{Random graphs with fixed degree sequence}
\label{secdegrees}

When $N_{i,j} = 1$ for all $(i,j)$, the support $\mathcal{S}_n$
reduces to the set $\mathcal{G}_n:= \{ 0,1\}^{n \choose2}$, which
encodes all undirected simple graphs on $n$ nodes: for any $x \in
\mathcal{G}_n$, the corresponding graph has an edge between nodes $i$
and $j$, with $i < j$, if and only if $x_{i,j} = 1$. In this case, the
$\beta$-model yields a class of distributions for random undirected
simple graphs on $n$ nodes, where the edges are mutually independent
Bernoulli random variables with probabilities of success $\{ p_{i,j}, i
< j\}$ satisfying (\ref{eqpij}). Then, by (\ref{eqsuffstat}), the
$i$th minimal sufficient statistic $d_i$ is the \textit{degree} of node
$i$, that is, the number of nodes adjacent to $i$, and the vector
$d(x)$ of sufficient statistics is the \textit{degree sequence} of the
observed graph $x$. This is the version of the $\beta$-model studied
by \citet{CDS10}.

\section{\texorpdfstring{Existence of the MLE for the $\beta$-model}
{Existence of the MLE for the beta-model}}
\label{secexistence}

We now derive a necessary and sufficient condition for the existence of
the MLE of the natural parameter $\beta$ or, equivalently, of the
probability parameters $\{ p_{i,j}, i < j\}$ as defined in (\ref
{eqpij}). For a given $x \in\mathcal{S}_n$, we say that the MLE does
not exist when
\[
\Bigl\{ \beta^* \dvtx p_{\beta^*}(x) = \sup_{\beta\in\mathbb{R}^n}
p_\beta(x) \Bigr\} = \varnothing,
\]
where $p_\beta(x)$ is given in (\ref{eqdens}).
For the natural parameters, nonexistence of the MLE implies that we
cannot attain the supremum of the likelihood function (\ref{eqdens})
by any finite vector in $\mathbb{R}^n$. For the probability
parameters, nonexistence signifies that the supremum of (\ref{eqlik})
cannot be attained by any set of probability values bounded away from
$0$ and $1$, and satisfying the equations from (\ref{eqpij}). Either
way, nonexistence of the MLE implies that only a random subset of the
model parameters
is estimable; see \citet{MLE}.

Our analysis on the existence of the MLE and parameter estimability for
the $\beta$-model is based on a geometric object that plays a key role
throughout the rest of the paper: the \textit{polytope of degree
sequences}.
To this end, we note that, for each $x \in\mathcal{S}_n$, we can
obtain the vector of sufficient statistics $d(x)$ for the $\beta
$-model as
\[
d(x) = \mathrm{A} x,
\]
where $\mathrm{A}$ is the $n \times{n \choose2}$ design matrix equal to
the node-edge incidence matrix of a complete graph on $n$ nodes.
Specifically, we index the rows of $\mathrm{A}$ by the node labels $i
\in\{
1,\ldots,n\}$, and the columns by the set of all pairs $(i,j)$ with
$i<j$, ordered lexicographically.
The entries of $\mathrm{A}$ are ones along the coordinates $(i,(i,j))$ and
$(j,(i,j))$ for $i<j$, and zeros otherwise.
For instance, when $n=4$
\[
\mathrm{A} = \lleft[ %
\matrix{ 1 & 1 & 1 & 0 & 0 & 0
\cr
1 & 0 & 0 & 1 & 1 & 0
\cr
0 & 1 & 0 & 1 & 0 & 1
\cr
0 & 0 & 1 & 0 & 1 & 1} \rright],
\]
where we index the columns lexicographically by the pairs $(1,2)$,
$(1,3)$, $(1,4)$, $(2,3)$, $(2,4)$ and $(3,4)$. In particular, for any
undirected simple graph $x \in\mathcal{G}_n$, $\mathrm{A} x$ is the
associated degree sequence.
The polytope of degree sequences $P_n$ is the convex hull of all
possible degree sequences, that is,
\[
P_n: = \operatorname{convhull} \bigl( \{ \mathrm{A} x, x \in
\mathcal{G}_n \} \bigr).
\]
The integral polytope $P_n$ is a well-studied object in graph theory;
for example, see Chapter 3 in \citet{MP95}. In particular, when $n=2$,
$P_n$ is
just a line segment in $\mathbb{R}^2$ connecting the points $(0,0)$
and $(1,1)$, while, for all $n \geq3$, $\operatorname{dim}(P_n) = n$.

We now fully characterize the existence of the MLE for the $\beta
$-model using the polytope of degree sequences in the following fashion.
For any $x \in\mathcal{S}_n$, let
\[
\tilde{p}_{i,j}: = \frac{x_{i,j}}{N_{i,j}},\qquad i < j,
\]
and set $\tilde{d} = \tilde{d}(x) \in\mathbb{R}^n$ to be the vector
with coordinates
%
%
\begin{equation}
\label{eqdtilde} \tilde{d}_i: = \sum_{j < i}
\tilde{p}_{j,i} + \sum_{j > i} \tilde
{p}_{i,j},\qquad i=1,\ldots,n,
\end{equation}
a rescaled version of the sufficient statistics (\ref{eqsuffstat}),
normalized by the number of observations. In particular, for the random
graph model, $\tilde{d} = d$.

%
\begin{theorem}\label{thmmle}
Let $x \in\mathcal{S}_n$ be the observed vector of edge counts. The
MLE exists if and only if $\tilde{d}(x) \in\operatorname{int}(P_n)$.
\end{theorem}

Theorem~\ref{thmmle} verifies the conjecture contained in Addendum A
in \citet{CDS10} for the random graph model:\vadjust{\goodbreak} the MLE exists if and
only if the degree sequence belongs to the interior of $P_n$. This
result follows from the standard properties of exponential families;
see Theorem~9.13 in \citet{BN78} or Theorem 5.5 in \citet{BROWN86}.
It also confirms the observation made by \citet{CDS10} that the MLE
never exists if $n=3$: indeed, since $P_3$ has exactly $8$ vertices, as
many as the possible graphs on $3$ nodes, no degree sequence can be
inside~$P_3$.

We conclude by taking note that, by representing the sufficient
statistics as a linear mapping $d = \mathrm{A} x$, we can recast the
$\beta
$-model as a log-linear model with design matrix $\mathrm{A}^\top$ and
product-multinomial scheme, with ${n \choose2}$ sampling constraints,
one for each edge. This simple yet far reaching observation allows us,
among the other things, to design algorithms for detecting nonexistence
of the MLE and identifying estimable parameters under a nonexistent
MLE, as explained in the supplementary material to this article.

\subsection{Parameter estimability under a nonexistent MLE}

The geometric nature of Theorem~\ref{thmmle} has important
consequences. First, it allows us to identify the patterns of observed
edge counts that cause nonexistence of the MLE; that is, the sample
points for which the MLE is undefined. Second, it yields a complete
description of estimability of the edge probability parameters under a
nonexistent MLE, a key issue for correct evaluation of degrees of
freedom of the model. The next result addresses the last two points.

%
\begin{lemma}\label{corfacial}
A point $y$ belongs to the interior of some face $F$ of $P_n$ if and
only if there exists a set $\mathcal{F} \subset\{ (i,j), i < j\}$
such that
%
%
\begin{equation}
\label{eqAp} y= \mathrm{A} p,
\end{equation}
where $p = \{ p_{i,j} \dvtx i < j, p_{i,j} \in[0,1] \} \in\mathbb
{R}^{{n \choose2}}$ is such that $p_{i,j} \in\{0,1\}$ if $(i,j)
\notin\mathcal{F}$ and $p_{i,j} \in(0,1)$ if $(i,j) \in\mathcal{F}$.
The set $\mathcal{F}$ is uniquely determined by the face $F$ and is the
maximal set for which (\ref{eqAp}) holds.
\end{lemma}

Following \citet{gms06} and \citet{MLE}, we refer to any such set
$\mathcal{F}$ a \textit{facial set} of $P_n$ and its complement,
$\mathcal{F}^c = \{ (i,j) \dvtx i < j\} \setminus\mathcal{F}$, a
\textit{co-facial set}. Facial sets form a lattice that is isomorphic
to the
face lattice of $P_n$ [\citet{MLE}, Lemma~5]. Thus the faces of $P_n$
are in one-to-one correspondence with the facial sets of $P_n$ and, for
any pair of faces $F$ and $F'$ of $P_n$ with associated facial sets
$\mathcal{F}$ and $\mathcal{F}'$, $F \cap F' = \varnothing$ if and only
if $\mathcal{F} \cap\mathcal{F}' = \varnothing$ and $F \subset F'$ if
and only if $\mathcal{F} \subset\mathcal{F}'$. In details, for a point
$x \in\mathcal{S}_n$, $d(x) = \mathrm{A} x$ belongs to the interior of a
face $F$ of $P_n$ if and only if there exists a nonnegative $p$ such
that $d(x) = \mathrm{A} p$, where $\mathcal{F} = \{ (i,j) \dvtx0 < p_{i,j}
< 1\}$ is the facial set corresponding to $F$. By the same token, $y
\in\operatorname{int}(P_n)$ if and only if $y = \mathrm{A} p$ for a
vector $p$ with
coordinates strictly between $0$ and $1$.

Facial sets have statistical relevance for two reasons.
First, nonexistence of the MLE can be described combinatorially in
terms of co-facial sets, that is, patterns of edge counts that are
either $0$ or $N_{i,j}$. In particular, the MLE does not exist if and
only if the set $\{ (i,j) \dvtx i < j, x_{i,j} =
0$ or $N_{i,j}\}$ contains a co-facial set.
Second, apart from exhausting all possible patterns of forbidden
entries in the table leading to a nonexistent MLE, facial sets specify
which probability parameters are estimable.
In fact, inspection of the likelihood function (\ref{eqlik}) reveals
that, for any observable set of counts $\{x_{i,j} \dvtx i < j \}$,
there always exists a unique maximizer $\widehat{p} = \{ \widehat
{p}_{i,j}, i < j \}$ which, by strict concavity, is uniquely determined
by the first order optimality conditions
\[
\tilde{d}(x) = \mathrm{A} \widehat{p},
\]
also known as the moment equations. Existence of the MLE is then
equivalent to $0 < \widehat{p}_{i,j} < 1$ for all $i<j$. When the MLE
does not exist, that is, when $\tilde{d}$ is on the boundary of $P_n$,
the moment equations still hold, but the entries of the optimizer $\{
\widehat{p}_{i,j}, i < j \}$, known as the \textit{extended MLE}, are no
longer strictly between $0$ and~$1$. Instead, by Lemma
\ref{corfacial}, the extended MLE is such that $\widehat{p}_{i,j} =
\tilde{p}_{i,j} \in\{0,1\}$ for all $(i,j) \in\mathcal{F}^c$.
Furthermore, it is possible to show [see, e.g., \citet{morton2008}]
that $\widehat{p}_{i,j} \in(0,1)$ for all $(i,j) \in\mathcal{F}$.
Therefore, when the MLE does not exist, only the probabilities $\{
p_{i,j}, (i,j) \in\mathcal{F} \}$ are estimable by the extended MLE.
We refer the reader to \citet{BN78}, \citet{BROWN86},
\citet{MLE} and
references therein, for details about the theory of extended
exponential families and extended maximum likelihood estimation in
log-linear models.

To summarize, while co-facial sets encode the patterns of table entries
leading to a nonexistent MLE, facial sets indicate which probability
parameters are estimable. A similar, though more involved
interpretation holds for the estimability of the natural parameters,
for which the reader is referred to \citet{MLE}.
Further, for a given sample point $x$, the realized facial set and its
cardinality are both random, as they depend on the actual value of the
observed sufficient statistics $\mathrm{A} x$. This implies that, with a
nonexistent MLE, the set of estimable parameters is itself random.

\section{The boundary of $P_n$}
\label{secapplic}

Theorem~\ref{thmmle} and Lemma~\ref{corfacial} show that the
boundary of the polytope $P_n$ plays a fundamental role in determining
the existence of the MLE for the $\beta$-model and in specifying which
parameters are estimable. In particular, the larger the number of faces
(i.e., facial sets) of $P_n$ the higher the complexity of the $\beta
$-model as measured by the numbers of possible patterns of edge counts
for which the MLE does not exist. Therefore, gaining an even basic
understanding of~the number and of the types of co-facial patterns will
provide valuable insights into the behavior of the $\beta$-model.
Below we further elaborate on the consequences of the results
established in Section~\ref{secexistence} and present a small
selection of examples of co-facial sets associated to the facets of $P_n$.

Though the discussion and examples of this section will reveal a number
of subtle issues, we believe that the key message is two-fold. First,
the combinatorial complexity of $P_n$, measured by both the number of
the types of co-facial sets, grows very fast with $n$, with the
co-facial sets associated to node degrees bounded away from $0$ and
$n-1$ vastly outnumbering the easily detectable cases of minimal or
maximal degree. Second, since complete enumeration of the faces of
$P_n$ is impractical, it is important to devise algorithms for
detecting a nonexistent MLE and identifying the facial sets of
estimable parameters. Both these issues become more severe in large and
sparse networks, where it is expected that the exploding number of
possible nontrivial co-facial set renders estimation of the model
parameters more difficult. Later in Section~\ref{secasymptotics}, we
will derive conditions, based on the geometry of $P_n$ that prevents
this from happening, with large probability for large $n$.

\subsection{The combinatorial complexity of $P_n$}
Mahadev and Peled (\citeyear{MP95}) describe the facet-defining inequalities of $P_n$, for
all $n \geq4$ (when $n \leq3$ the problem is of little interest), a
result we use later in Section~\ref{secasymptotics}. Let $\mathcal
{P}$ be the set of all pairs $(S,T)$ of disjoint nonempty subsets of
$\{1,\ldots,n\}$, such that $|S \cup T| \in\{ 2,\ldots, n-3,n\}$.
For any $(S,T) \in\mathcal{P}$ and $y \in P_n$, let
%
%
\begin{equation}
\label{eqg} g(S,T,y,n):= |S|\bigl( n - 1 - |T|\bigr) - \sum_{i \in S}
y_i + \sum_{i \in T} y_i.
\end{equation}

%
\begin{theorem}[{[Theorem 3.3.17 in \citet{MP95}]}]\label{thmMP}\label{theoremMP}
Let $n \geq4$ and $y\in P_n$. The facet-defining inequalities of $P_n$
are:
\begin{longlist}[(iii)]
\item[(i)] $ y_i \geq0$, for $i=1,\ldots,n$;
\item[(ii)] $y_i \leq n-1$, for $i=1,\ldots,n$;
\item[(iii)] $g(S,T,y,n) \geq0$, for all $(S,T) \in\mathcal{P}$.
\end{longlist}
\end{theorem}

Even with the exhaustive characterization of $P_n$ provided by
Theorem~\ref{theoremMP}, understanding the combinatorial complexity
of $P_n$ (i.e., the collection of all its faces and their inclusion
relations) is far from trivial. \citet{ST91} studied the number faces
of the polytope of degree sequences $P_n$ and derived an expression for
computing the entries of the $f$-vector of $P_n$. The $f$-vector of an
$n$-dimensional polytope is the vector of length $n$ whose $i$th entry
contains the number of $i$-dimensional faces, $i=0,\ldots,n-1$. For
example, the $f$-vector of $P_8$ is the $8$-dimensional vector
\[
(334\mbox{,}982,1\mbox{,}726\mbox{,}648,3\mbox{,}529\mbox{,}344,
3\mbox{,}679\mbox{,}872,2\mbox{,}074\mbox{,}660,610\mbox{,}288,81\mbox
{,}144,3322).
\]
Thus, $P_8$ is an $8$-dimensional polytope with $334\mbox{,}982$ vertices,
$1\mbox{,}726\mbox{,}648$ edges and so on, up to $3322$ facets. Also,
according to
Stanley's formula, the number of facets of $P_4$, $P_5$, $P_6$ and
$P_7$ are $22$, $60$, $224$ and $882$, respectively [these numbers
correspond to the numbers we obtained with the software \texttt{polymake},
using the methods described in the supplementary material to this
article; see \citet{polymake}]. Stanley's analysis showed that the combinatorial complexity
of $P_n$ is extraordinarily large, with both the number of vertices,
and the number of facets growing at least exponentially in~$n$, and
consequently, the tasks of identifying points on the boundary of $P_n$
and the associated facial set are far from trivial. For instance,
computing directly the number of vertices of $P_{10}$ is prohibitively
expensive, even using one of the best known algorithms, such as the one
implemented in the software \texttt{minksum}; see \citet{minksum}. To
overcome these problems we have devised an algorithm for detecting
boundary points and the associated facial sets that can handle networks
with up to hundreds of nodes. We report on this algorithm, which is
based on a log-linear model reparametrization and is equivalent to what
is known in computational geometry as the ``Cayley trick,'' in the
supplementary material. Using the methods described there, we were
able to identify a few interesting cases in which the MLE does not
exist, most of which have gone unrecognized in the statistical
literature. Below we describe some of our computations for the purpose
of elucidating the results derived in Section~\ref{secexistence}.

\subsection{Some examples of co-facial sets} Recall that we can
represent the data as a $n \times n$ table of counts with structural
zero diagonal elements and where the $(i,j)$th entry of the table
indicates the number of times, out of $N_{i,j}$, in which we observed
the edges $(i,j)$. In our examples, empty cells correspond to facial
sets and may contain arbitrary count values, in contrast to the cells
in the co-facial sets that contain either a zero value or a maximal
value, namely $N_{i,j}$. Lemma~\ref{corfacial} implies that extreme
count values of this nature are precisely what leads to the
nonexistence of the MLE.
The pattern shown on the left of Table~\ref{tabtab1} provides an
instance of a co-facial set, which corresponds to a facet of $P_4$.
Assume for simplicity that the empty cells contain counts bounded away
%
%
\begin{table}[b]
\tablewidth=280pt
\caption{Left: co-facial set leading to a nonexistent MLE. Center: an
example of data exhibiting the pattern of counts consistent with the
co-facial set on the left when $N_{i,j} = 3$ for all $i \neq j$. Right:
table of the extended MLE of the estimated probabilities}
\label{tabtab1}
\begin{tabular*}{\tablewidth}{@{\extracolsep{\fill}}ccc@{}}
\begin{tabular}{@{}cccc@{}}
\hline
$\times$ & 0 & $ $ & $ $\\
{$N_{1,2}$} & $\times$ & $ $ & $ $\\
$ $ & $ $ & $\times$ & {$N_{3,4}$}\\
$ $&$ $ & 0 & $\times$\\
\hline
\end{tabular}
&
\begin{tabular}{@{}cccc@{}}
\hline
$\times$ & 0 & 1 & 2\\
3 & $\times$ & 2 & 1\\
2 & 1 & $\times$ & 3\\
1 & 2 & 0 &$\times$\\
\hline
\end{tabular}
&
\begin{tabular}{@{}cccc@{}}
\hline
$\times$ & 0 & 0.5 & 0.5\\
1 &$\times$ & 0.5 & 0.5\\
0.5 & 0.5 & $\times$ & 1\\
0.5 & 0.5 & 0 & $\times$\\
\hline
\end{tabular}
\end{tabular*}
\end{table}
from 0 and $N_{i,j}$. Then the sufficient statistics $\tilde{d}$ are
also bounded away from $0$ and $n-1$, and so are the row and column
sums of the normalized counts $\{ \frac{x_{i,j}}{N_{i,j}} \dvtx i
\neq j\}$, yet the MLE does not exist.
This is further illustrated in Table~\ref{tabtab1}, center, which
shows an instance of data with $N_{i,j} =3$ for all $i \neq j$,
satisfying the above pattern and, on the right, the probability values
%
%
\begin{table}
\tablewidth=280pt
\caption{Examples of a co-facial set leading to a nonexistent
MLE.\break
Left: $\tilde{d}_2 = 0$. Right: example where the degrees are all
bounded away from $0$ and $3$, the MLE does not exist}
\label{tabtab3}
\begin{tabular*}{\tablewidth}{@{\extracolsep{\fill}}cccccc@{}}
&&
\begin{tabular}{@{}cccc@{}}
\hline
$\times$ & 0 &  & \\
{$N_{1,2}$} & $\times$ & 0 & 0\\
 & {$N_{3,2}$} & $\times$ & \\
 & {$N_{4,2}$} &  & $\times$ \\
\hline
\end{tabular}
&
\begin{tabular}{@{}cccc@{}}
\hline
$\times$ & 0 &  & 0\\
{$N_{1,2}$} & $\times$ &  & 0\\
 &  & $\times$ & \\
{$N_{4,1}$} & {$N_{4,2}$} &  & $\times$ \\
\hline
\end{tabular}
&&
\end{tabular*}
\end{table}
maximizing the log-likelihood function.
Notice that, because the MLE does not exist, the supremum of the
log-likelihood under the natural parametrization is attained in the
limit by any sequence of natural parameters $\{ \beta^{(k)}\}$ of the
form $\beta^{(k)} = (-c_k,-c_k,c_k,c_k)$, where $c_k \rightarrow
\infty$ as $k \rightarrow\infty$.
As a result, some of these probability values are $0$ and $1$. The
order of the pattern is crucial.
In Table~\ref{tabtab3} we show, on the left, another example of a
co-facial set that is easy to detect, since it corresponds to a value
of $0$ for the normalized sufficient statistic $\tilde{d}_2$. Indeed,
from cases (i) and (ii) of Theorem~\ref{thmMP}, the MLE
does not exist if $\tilde{d}_i = 0$ or $\tilde{d}_i = n-1$, for some
$i$. On the right, we show a co-facial set that is instead compatible
with normalized sufficient statistics being bounded away from $0$ and
$n-1$. Finally, in Table~\ref{taballcases} we list all 22 co-facial
sets associated with the facets of~$P_4$, including the cases already shown.

%
\begin{table}
\caption{All possible co-facial sets for $P_4$ corresponding to the
facets of $P_4$\break (empty cells indicate arbitrary entry values)}
\label{taballcases}
\begin{tabular}{@{}c@{\qquad}c@{\qquad}c@{}}
\hline
\begin{tabular*}{98pt}{@{\extracolsep{4in minus 4in}}cccc@{}}
$\times$ & 0 & $ $ & $ $\\
{\bf$N_{1,2}$} & $\times$ & $ $ & $ $\\
$ $ & $ $ & $\times$ & {\bf$N_{3,4}$}\\
$ $&$ $ & 0 & $\times$\\
\end{tabular*}
&
\begin{tabular*}{98pt}{@{\extracolsep{4in minus 4in}}cccc@{}}
$\times$ & 0 & $ $ & $ $\\
{\bf$N_{1,2}$} & $\times$ & 0 & 0\\
$ $ & {\bf$N_{3,2}$} & $\times$ & $ $\\
$ $ & {\bf$N_{4,2}$} & $ $ & $\times$ \\
\end{tabular*}
&
\begin{tabular*}{98pt}{@{\extracolsep{4in minus 4in}}cccc@{}}
$\times$ & 0 & $ $ & 0\\
{\bf$N_{1,2}$} & $\times$ & $ $ & 0\\
$ $ & $ $ & $\times$ & $ $\\
{\bf$N_{4,1}$} & {\bf$N_{4,2}$} & $ $ & $\times$ \\
\end{tabular*}
\\
\begin{tabular*}{98pt}{@{\extracolsep{4in minus 4in}}cccc@{}}
\hline
$\times$ & 0 & 0 & 0\\
{\bf$N_{1,2}$} & $\times$ & $ $ & \\
$N_{1,3}$ & $ $ & $\times$ & $ $\\
{\bf$N_{4,1}$} & $ $ & $ $ & $\times$ \\
\end{tabular*}
&
\begin{tabular*}{98pt}{@{\extracolsep{4in minus 4in}}cccc@{}}
\hline
$\times$ & 0 & 0 & $ $\\
{\bf$N_{1,2}$} & $\times$ & 0 & \\
$N_{1,3}$ & $N_{2,3}$ & $\times$ & $ $\\
& $ $ & $ $ & $\times$ \\
\end{tabular*}
&
\begin{tabular*}{98pt}{@{\extracolsep{4in minus 4in}}cccc@{}}
\hline
$\times$ & $ $ & 0 & 0\\
$ $ & $\times$ & $ $ & $ $\\
$N_{1,3}$ & $ $ & $\times$ & 0\\
$N_{1,4}$ & $ $ & $N_{3,4}$ & $\times$ \\
\end{tabular*}
\\
\begin{tabular*}{98pt}{@{\extracolsep{4in minus 4in}}cccc@{}}
\hline
$\times$ & $ $ & 0 & $ $\\
$ $ & $\times$ & 0 & $ $\\
$N_{1,3}$ & $N_{2,3}$ & $\times$ & 0\\
& & $N_{3,4}$ & $\times$ \\
\end{tabular*}
&
\begin{tabular*}{98pt}{@{\extracolsep{4in minus 4in}}cccc@{}}
\hline
$\times$ & $N_{1,2}$ & $N_{1,3}$ & $N_{1,4}$\\
$ $ 0 $ $ & $\times$ & & $ $\\
0 & $ $ & $\times$ & $ $\\
0 & & & $\times$ \\
\end{tabular*}
&
\begin{tabular*}{98pt}{@{\extracolsep{4in minus 4in}}cccc@{}}
\hline
$\times$ & & $N_{1,3}$ & $N_{1,4}$\\
$ $ & $\times$ & & $ $\\
0 & $ $ & $\times$ & $N_{3,4}$\\
0 & & 0 & $\times$ \\
\end{tabular*}
\\
\begin{tabular*}{98pt}{@{\extracolsep{4in minus 4in}}cccc@{}}
\hline
$\times$ & $N_{1,2}$ & $N_{1,3}$ & $ $\\
$ $ 0 $ $ & $\times$ & $N_{2,3}$ & $ $\\
0 & 0 & $\times$ & \\
& & $ $ & $\times$ \\
\end{tabular*}
&
\begin{tabular*}{98pt}{@{\extracolsep{4in minus 4in}}cccc@{}}
\hline
$\times$ & $ $ & $N_{1,3}$ & $ $\\
$ $ & $\times$ & $N_{2,3}$ & $ $\\
0& 0 & $\times$ & $N_{3,4}$ \\
& &0 & $\times$ \\
\end{tabular*}
&
\begin{tabular*}{98pt}{@{\extracolsep{4in minus 4in}}cccc@{}}
\hline
$\times$ & $N_{1,2}$ & $ $ & $N_{1,4}$\\
0 & $\times$ & $ $ & $N_{2,4}$ \\
$ $& $ $ & $\times$ & $ $\\
0 & 0 & & $\times$ \\
\end{tabular*}
\\
\begin{tabular*}{98pt}{@{\extracolsep{4in minus 4in}}cccc@{}}
\hline
$\times$ & $ $ & $ $ & $N_{1,4}$\\
$ $ & $\times$ & $ $ & $N_{2,4}$\\
& & $\times$ & $N_{3,4}$\\
0& 0& 0& $\times$ \\
\end{tabular*}
&
\begin{tabular*}{98pt}{@{\extracolsep{4in minus 4in}}cccc@{}}
\hline
$\times$ & $ $ & 0 & $ $\\
$ $ & $\times$ & 0 & $ $\\
$N_{1,3}$ & $N_{2,3}$ & $\times$ & 0 \\
& & $N_{3,4}$ & $\times$ \\
\end{tabular*}
&
\begin{tabular*}{98pt}{@{\extracolsep{4in minus 4in}}cccc@{}}
\hline
$\times$ & $N_{1,2}$ & $ $ & $ $\\
0 & $\times$ & 0 & 0 \\
$ $ & $N_{2,3}$ & $\times$ & $ $\\
$ $ & $N_{2,4}$ & $ $ & $\times$ \\
\end{tabular*}\\
\begin{tabular*}{98pt}{@{\extracolsep{4in minus 4in}}cccc@{}}
\hline
$\times$ & $ $ & $ $ & $ $\\
$ $ & $\times$ & 0 & 0\\
$ $ & $N_{2,3}$ & $\times$ & 0\\
& $N_{2,4}$ & $N_{3,4}$ & $\times$ \\
\end{tabular*}
&
\begin{tabular*}{98pt}{@{\extracolsep{4in minus 4in}}cccc@{}}
\hline
$\times$ & $ $ & $ $ & $ $ 0 $ $\\
$ $ & $\times$ & $ $ & $ $ 0 $ $\\
& & $\times$ & $ $ 0 $ $ \\
$N_{1,4}$ & $N_{2,4}$ & $N_{3,4}$ & $\times$ \\
\end{tabular*}
&
\begin{tabular*}{98pt}{@{\extracolsep{4in minus 4in}}cccc@{}}
\hline
$\times$ & $N_{1,2}$ & $ $ & $ $\\
0 & $\times$ & $ $ & $ $\\
$ $ & $ $ & $\times$ & 0\\
& $ $& $N_{3,4}$ & $\times$ \\
\end{tabular*}
\\
\begin{tabular*}{98pt}{@{\extracolsep{4in minus 4in}}cccc@{}}
\hline
$\times$ & $ $ & 0 & $ $\\
$ $ & $\times$ & $ $ & $N_{2,4}$\\
$N_{1,3}$ & $ $ & $\times$ & $ $ \\
$ $ & 0 & $ $ & $\times$ \\
\hline
\end{tabular*}
&
\begin{tabular*}{98pt}{@{\extracolsep{4in minus 4in}}cccc@{}}
\hline
$\times$ & $ $ & $N_{1,3}$ & $ $\\
$ $ & $\times$ & $ $ & 0\\
0 & $ $ & $\times$ & $ $ \\
$ $ & $N_{2,4}$ & $ $ & $\times$ \\
\hline
\end{tabular*}
&
\begin{tabular*}{98pt}{@{\extracolsep{4in minus 4in}}cccc@{}}
\hline
$\times$ & $ $ & $ $ & $N_{1,4}$\\
$ $ & $\times$ & 0 & $ $\\
$ $ & $N_{2,3}$ & $\times$ & $ $ \\
0 & $ $& $ $ & $\times$ \\
\hline
\end{tabular*}
\\
&\begin{tabular*}{98pt}{@{\extracolsep{4in minus 4in}}cccc@{}}
$\times$ & $ $ & $ $ & 0\\
$ $ & $\times$ & $N_{2,3}$ & $ $\\
$ $ & 0 & $\times$ & $ $ \\
$N_{1,4}$ & $ $& $ $ & $\times$ \\
\end{tabular*}
&\\
\hline
\end{tabular}
\end{table}

In general, there are $2n$ facets of $P_n$ that are determined by one
$\tilde{d}_i$ equal to $0$ or $n-1$.
Thus, just by inspecting the row sums or the observed sufficient
statistics, we can detect only $2n$ co-facial sets associated to as
many facets of $P_n$. Comparing this number to the entries of the
$f$-vector calculated in \citet{ST91}, however,
and as our computations confirm, most of the facets of $P_n$ do not
yield co-facial sets of this form. Since the number of facets appears
to grow exponentially in $n$, we conclude that most of the co-facial
sets do not appear to arise in this fashion. Thus, at least
combinatorially, patterns of data counts leading to the nonexistence of
MLEs but with the normalized degree bounded away from $0$ and $n-1$ are
much more frequent, especially in larger networks.

\subsection{The random graph case}

In the special case of $N_{i,j} = 1$ for all $i<j$, which is equivalent
to a model for random undirected graphs, points on the boundary of
$P_n$ are, by construction, degree sequences and have a direct
graph-theoretical interpretation. We say that a subset of a set of
nodes of a given graph is \textit{stable} if it induces a subgraph with no
edges and a \textit{clique} if it induces a complete subgraph.

%
\begin{lemma}[{[Lemma 3.3.13 in \citet{MP95}]}]\label{lemMP}
Let $d$ be a degree sequence of a graph $\mathcal G$
that lies on the boundary of $P_n$. Then either $d_i = 0$, or $d_i =
n-1$ for some $i$, or there exist nonempty and disjoint subsets $S$
and $T$ of $\{ 1,\ldots,n\}$ such that:
\begin{enumerate}[(4)]
\item[(1)]$S$ is clique of $\mathcal{G}$;
\item[(2)] $T$ is a stable set of $\mathcal{G}$;
\item[(3)] every vertex in $S$ is adjacent to every vertex in $(S \cup
T)^c$ in $\mathcal{G}$;
\item[(4)] no vertex of $T$ is adjacent to any vertex of $(S \cup
T)^c$ in $\mathcal{G}$.
\end{enumerate}
\end{lemma}

\begin{figure}

\includegraphics{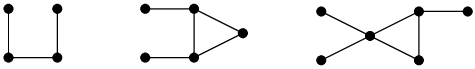}

\caption{Examples of random graphs on $4$ (left), $5$ (center) and $6$
(right) nodes with node degrees bounded away from $0$ and $n-1$ and for
which the MLE is not defined. Lemma~\protect\ref{lemMP} applies with
$S = \{
3,4\}$ and $T = \{1,2\}$ (left), with $S = \{ 2,3,4\}$ and $T =
\{1,5\}$ (center) and with $S = \{ 1,2,6\}$ and $T = \{3,4,5\}$
(right).} \label{tabtab5}
\end{figure}

Using Lemma~\ref{lemMP}, we can create virtually any example of a
random graph whose node degree sequence lies on the boundary of $P_n$.
In particular, we note that having node degrees bounded away from $0$
and $n-1$ \textit{is not} a sufficient condition for the existence of the
MLE, although its violation implies nonexistence of the MLE; see the
examples of Figure~\ref{tabtab5}. Nonetheless, Lemma~\ref{lemMP} is
of little or no practical use when it comes to detecting boundary
points and the associated co-facial sets, since checking for the
existence of a pair $(S,T)$ of subsets of nodes satisfying conditions
(1) through (4) is algorithmically impractical. In the
supplementary material to this article, we describe alternative
procedures that can be used in large networks.

%

Figure~\ref{tabtab5} shows three examples of graphs on $4$, $5$ and
$6$ nodes for which the MLE of the $\beta$-model is undefined even
though the node degrees are bounded away from $0$ and $n-1$ in all
cases. All the examples were constructed using directly Lemma \ref
{lemMP}, as explained in the caption. To the best of our knowledge,
even these very small examples of nonexistent MLEs are unknown to
practitioners and no available software for fitting the MLE is able to
detect nonexistence, much less identify the relevant facial set.

For the case $n=4$, our computations show that there are $14$ distinct
co-facial sets associated to the facets of $P_n$. Eight of them
correspond to degree sequences containing a $0$ or a $3$, and the
remaining six are shown in Table~\ref{tabtab4}, which we computed
numerically using the procedure described in the supplementary
%
%
\begin{table}
\caption{Patterns of zeros and ones yielding random graphs with
nonexistent MLE\break (empty cells indicate that the entry could be a $0$ or
a $1$)}
\label{tabtab4}
\begin{tabular}{@{}ccc@{}}
\begin{tabular}{@{}cccc@{}}
\hline
$\times$ & 0 & $ $ & $ $\\
1 & $\times$ & $ $ & $ $\\
$ $ & $ $ & $\times$ & 1\\
$ $ & $ $ & 0 & $\times$\\
\end{tabular}
&
\begin{tabular}{@{}cccc@{}}
\hline
$\times$ & $ $ & 0 & $ $\\
$ $ & $\times$ & $ $ & $ $\\
1 & $ $ & $\times$ & 1\\
$ $ & &0 & $\times$\\
\end{tabular}
&
\begin{tabular}{@{}cccc@{}}
\hline
$\times$ & $ $ & $ $ & 1\\
$ $ & $\times$ & 0 & $ $\\
$ $ & 1 & $\times$ & $ $\\
0 & $ $ &$ $ & $\times$\\
\end{tabular}
\\
\begin{tabular}{@{}cccc@{}}
\hline
$\times$ & 1 & $ $ & $ $\\
0 & $\times$ & $ $ & $ $\\
$ $ & $ $ & $\times$ & 0\\
$ $ & $ $ & 1 & $\times$\\
\hline
\end{tabular}
&
\begin{tabular}{@{}cccc@{}}
\hline
$\times$ & $ $ & 1 & $ $\\
$ $ & $\times$ & $ $ & $ $\\
0 & $ $ & $\times$ & 0\\
$ $ & & 1 & $\times$\\
\hline
\end{tabular}
&
\begin{tabular}{@{}cccc@{}}
\hline
$\times$ & $ $ & $ $ & 0\\
$ $ & $\times$ & 1 & $ $\\
$ $ & 0 & $\times$ & $ $\\
1 & $ $ &$ $ & $\times$\\
\hline
\end{tabular}
\end{tabular}
\end{table}
material. Notice that the three tables on the second row are obtained
from the first three tables by switching zeros with ones. Furthermore,
the number of the co-facial sets we found is smaller than the number of
facets of $P_n$, which is $22$, as shown in Table~\ref{taballcases}.
This is a consequence of the fact that the only observed counts in the
random graph model are $0$'s or $1$'s: it is in fact easy to see in Table
\ref{taballcases} that any co-facial set containing three zero counts
and three maximal counts $N_{i,j}$ is equivalent, in the random graph
case, to a node having degree zero or $3$.
However, as soon as $N_{i,j} \geq2$, the number of possible co-facial
sets matches the number of faces of $P_n$. Therefore, the condition
$N_{i,j} = 1$ is not inconsequential, as it appears to reduce the
numbers of observable patterns leading to a nonexistent MLE, though we
do not know the extent of the impact of such reduction in general.

\section{Existence of the MLE: Finite sample bounds}
\label{secasymptotics}

In this section we exploit the geometry of the boundary of $P_n$ from
Lemma~\ref{lemMP} to derive sufficient conditions that imply the
existence of the MLE with large probability as the size of the network
$n$ grows. These conditions essentially guarantee that the probability
of observing any of the super-exponentially many (in $n$) co-facial
sets of $P_n$ is polynomially small in $n$. Unlike in previous
analyses, our result does not require the network to be dense.

We make the simplifying assumption that $N_{i,j} = N$, for all $i$ and
$j$, where $N = N(n) \geq1$ could itself depend on $n$.
Recall the random vector $\tilde{d}$, whose coordinates are given in
(\ref{eqdtilde}) and let $\overline{d} = \mathbb{E}[\tilde{d}] \in
\mathbb{R}^n$ be its expected value under the $\beta$-model. Then
\[
\overline{d}_i = \sum_{j < i}
p_{j,i} + \sum_{j > i} p_{i,j},\qquad i=1,\ldots,n.
\]
We formulate sufficient conditions for the existence of the MLE in
terms of the entries of the vector $\overline{d}$.

%
\begin{theorem}\label{thmus}
Assume that, for all $n \geq\max\{ 4, 2 \sqrt{ c \frac{n \log
n}{N}} + 1 \}$, the vector $\overline{d}$ satisfies the conditions:
\begin{longlist}[(ii)]
\item[(i)] $\min_{i} \min\{ \overline{d}_i, n-1-\overline
{d}_i \} \geq2 \sqrt{ c \frac{n \log n}{N}} + C$,
\item[(ii)] $\min_{(S,T) \in\mathcal{P}} g(S,T,\overline{d},n) >|S
\cup T| \sqrt{ c \frac{n \log n}{N}} + C$,
\end{longlist}
where $c>1/2$ and $C \in(0, \frac{n-1}{2} - \sqrt{ c \frac{n
\log n}{N}} )$. Then, with probability at least $1 - \frac
{2}{n^{2c-1}}$, the MLE exists.
\end{theorem}

When $N$ is constant, for example, when $N=1$ as in the random graph
case, we can relax the conditions of Theorem~\ref{thmus} by requiring
condition (ii) to hold only over subsets $S$ and $T$ of
cardinality of order $\Omega(\sqrt{n \log n})$. While we present this
result in greater generality by assuming only that $n \geq N$, we do
not expect it to be sharp in general when $N$ grows with $n$.

%
\begin{corollary}\label{corus}
Let $n \geq\max\{ N,4, 2 \sqrt{ c n \log n} + 1 \}$, $c>1$ and $C
\in(0, \frac{n-1}{2} - \sqrt{c n \log n} )$. Assume the
vector $\overline{d}$ satisfies the conditions:
\begin{longlist}[(ii$'$)]
\item[(i$'$)] $\min_{i} \min\{ \overline{d}_i, n-1-\overline
{d}_i \} \geq2 \sqrt{c n \log n} + C$,
\item[(ii$'$)] $\min_{(S,T) \in\mathcal{P}_n} g(S,T,\overline{d},n)
>|S \cup T| \sqrt{c n \log n} + C$,
\end{longlist}
where
\[
\mathcal{P}_n:= \bigl\{ (S,T) \in\mathcal{P} \dvtx\min\bigl\{ |S|,|T|\bigr\}
> \sqrt{c n \log n} + C\bigr\},
\]
where the set $\mathcal{P}$ was defined before Theorem~\ref{thmMP}.
Then the MLE exists with probability at least $1 - \frac
{2}{n^{2c-2}}$. If $N=1$, it is sufficient to have $c>1/2$, and the MLE
exists with probability larger than $1 - \frac{2}{n^{2c-1}}$.
\end{corollary}

\subsection*{Discussion and comparison with previous work}

Since $|S \cup T| \leq n$, one could replace assumption (ii) of
Theorem~\ref{thmus} with the simpler but stronger condition
\[
\min_{(S,T) \in\mathcal{P}_n} g(S,T,\overline{d},n) > n^{3/2}\sqrt{c
\log n} + C_n.
\]
Then, if we assume for simplicity that $N$ is a constant, as in
Corollary~\ref{corus}, the MLE exists with probability tending to one
at a rate that is polynomial in $n$ whenever
\[
\min_{i} \min\{ \overline{d}_i, n-1-
\overline{d}_i \} = \Omega(\sqrt{ n \log n} )
\]
and, for all pairs $(S,T) \in\mathcal{P}$,
\[
g(S,T,\overline{d},n) >\Omega\bigl( n^{3/2} \sqrt{\log n} \bigr).
\]

For the case $N = 1$, we can compare Corollary~\ref{corus} with
Theorem 3.1 in \citet{CDS10}, which also provides sufficient
conditions for the existence of the MLE with probability no smaller
than $1 - \frac{1}{n^{2c -1}}$ (for all $n$ large enough). Their
result appears to be stronger than ours, but that is actually not the
case as we now explain. In fact, their conditions require that, for
some constant $c_1$, $c_2$ and $c_3$ in $(0,1)$, $c_1 (n-1) < d_i < c_2
(n-1)$ for all $i$ and
%
%
\begin{equation}
\label{eqSD2} |S|\bigl(|S|-1\bigr) - \sum_{i \in S}
d_i + \sum_{i \notin S} \min\bigl\{
d_i, |S| \bigr\} > c_3 n^2
\end{equation}
for all sets $S$ such that $|S| > (c_1)^2 n^2$.
For any nonempty subsets $S \subset\{ 1,\ldots,n\}$ and $T \subset
\{ 1,\ldots,n\} \setminus S$,
\[
\sum_{i \notin S} \min\bigl\{ d_i, |S| \bigr\} \leq\sum
_{i \in T} d_i + |S|\bigl|(S \cup
T)^c\bigr|,
\]
which implies that
\[
|S|\bigl(n-1-|T|\bigr) - \sum_{i \in S} d_i + \sum
_{i \in T} d_i > |S|\bigl(|S|-1\bigr) - \sum
_{i \in S} d_i + \sum_{i \notin S}
\min\bigl\{ d_i |S| \bigr\},
\]
where we have used the equality $n = |S| + |T| + |(S \cup T)^c|$. Thus
if (\ref{eqSD2}) holds for some nonempty $S \subset\{ 1,\ldots,n\}$,
it satisfies the facet conditions implied by all the pairs $(S,T)$, for
any nonempty set $T \subset\{ 1,\ldots,n\} \setminus S$. As a result,
for any subset $S$, condition (\ref{eqSD2}) is stronger than any of
the facet conditions of $P_n$ specified by $S$. In addition, we
weakened significantly the requirements in \citet{CDS10} that $c_1
(n-1) < d_i < c_2 (n-1)$ for all $i$ to $\min_{i} \min\{
\overline{d}_i, n-1-\overline{d}_i \} \geq2 \sqrt{c n \log n} +
C$. As a direct consequence of this weakening, we only need $|S| >
\sqrt{c n \log n} + C$ as opposed to $|S| > (c_1)^2 n^2$. Overall, in
our setting, the vector of expected degrees of the sequence of networks
is allowed to lie much closer to the boundary of $P_n$. As we explain
next, such weakening is significant, since the setting of \citet{CDS10}
only allows us to estimate an increasing number of probability
parameters (the edge probabilities) that are uniformly bounded away
from $0$ and $1$, while our assumptions allow for these probabilities
to become degenerate as the network size grows, and therefore hold even
in nondense network settings.

\subsubsection*{The nondegenerate case} We now briefly discuss the
case of sequences of networks for which $N=1$ and the edge
probabilities are uniformly bounded away from $0$ and $1$, that is,
%
%
\begin{equation}
\label{eqdense} \delta< p_{i,j} < 1 - \delta\qquad\forall i,j
\end{equation}
for some $\delta\in(0,1)$ independent of $n$. In this scenario, the
number of probability parameters to be estimated grows with $n$, but
their values are guaranteed to be nondegenerate.
It immediately follows from the nondegenerate assumption (\ref
{eqdense}) that $\overline{d} \in\operatorname{int}(P_n)$ and
%
%
\begin{equation}
\label{eqdidelta} \delta(n-1) < \overline{d}_i < (1 -\delta)
(n-1),\qquad i = 1,\ldots,n.
\end{equation}
Then, the same arguments we used in the proof of Corollary \ref
{corus} imply that the MLE exists with high probability. We provide a
sketch of the proof. First, we note that, with high probability,
$g(S,T,\tilde{d},n) \geq g(S,T,\overline{d},n) - |S \cup T| \Omega
( \sqrt{ n \log n} )$, for each pair $(S,T) \in\mathcal
{P}$. Furthermore, because of (\ref{eqdidelta}), it is enough to
consider only pairs $(S,T)$ of disjoint subsets of $\{1,\ldots,n\}$ of
sizes of order $\Omega(n)$. For each such pair,\vadjust{\goodbreak} the condition on
$\overline{d}_i$ further yields that $g(S,T,\overline{d},n)$ is of
order $\Omega(n^2)$, and, by Theorem~\ref{eqg} the MLE exists with
high probability.

In fact, the boundedness assumption of \citet{CDS10} that $\| \beta
\|
_\infty< L$ with $L$ independent of $n$, is equivalent to the
nondegenerate assumption (\ref{eqdense}), as we see from equation
(\ref{eqpij}).
Unlike \citet{CDS10}, who focus on the nondegenerate case, our
results hold under weaker scaling, as we only require, for instance, that
$\overline{d}_i$ be of order $\Omega( \sqrt{n \log n}
)$ for all $i$.
Relatedly, we note that the tameness condition of \citet{BH10} is
equivalent to $\delta< \widehat{p}_{i,j} < 1 - \delta$ for all $i$
and $j$ and a fixed $\delta\in(0,1)$, where $\widehat{p}_{i,j}$ is
the MLE of $p_{i,j}$. Therefore, the tameness condition is stronger
than the existence of the MLE. In fact, using again Theorem 1.3 in
\citet{CDS10}, for all $n$ sufficiently large, the tameness condition
is equivalent to the boundedness condition of \citet{CDS10}.

We conclude this section with two useful remarks. First, Theorem 1.3 in
\citet{CDS10} demonstrates\vspace*{2pt} that, when the MLE exists, ${\max_i} |
\widehat{\beta}_i - \beta_i | = O(\sqrt{ \log n / n})$, with
probability at least $1 - \frac{2}{n^{2c} - 1}$. Combined\vspace*{1pt} with our
Corollary~\ref{corus}, this implies that the MLE is a consistent
estimator under a growing network size and with edge probabilities
approaching the degenerate values of $0$ and $1$.

Second, after the submission of this article we learned about the
interesting asymptotic results of \citet{Y1,Y2}, who claim that, based
on a modification of the arguments of \citet{CDS10}, it is
possible to
show the MLE of the $\beta$-model exists and is uniformly consistent
if $L = o(\log n) $ and $L = o(\log\log n)$, respectively, where $L =
{\max_i} |\beta_i|$.

\section{Discussion and extensions}\label{secdiscussion}

We have used polyhedral geometry to analyze the conditions for
existence of the MLE of a generalized version of the $\beta$-model and
to derive finite sample bounds for the probability associated with the
existence of the MLE. Our results offer a novel and explicit
characterization of the patterns of edge counts leading to nonexistent
MLEs. The problem of nonexistence occurs in numbers and with a
complexity that was not previously known. Our results allow us to
sharpen conditions for existence of the MLE. Our analysis in
particular highlights the fact that requiring node degrees equal to $0$
and $n-1$ is only a sufficient condition for nonexistence of the MLE
and nonestimability of the edge probabilities. We show that we need
to account for many more edge patterns. We note that the use of
polyhedral geometry in statistical models for discrete data is a
hallmark of the theory of exponential families, but its considerable
potential for use and applications in the analysis of log-linear and
network models has only recently begun to be investigated;
\citet{MLE,ERGM}.

Our generalization of the $\beta$-model allows for Poisson and
binomial, not simply Bernoulli distributions for edges. Email databases
and others involving repeated transactions among pairs of parties
provides the simplest examples of situations for networks where edges
can occur multiple times. These are often analyzed as weighted networks
but that may not necessarily make as much sense as using a Poisson for
random numbers of occurrences.

As our results indicate, the nonexistence of the MLE is equivalent to
nonestimability of a subset of the parameters of the model, but by no
means does it imply that no statistical inference can take place. In
fact, when the MLE does not exist, there always exists a ``restricted''
$\beta$-model that is specified by the appropriate facial set, and for
which all parameters are estimable. Thus, for such a small model,
traditional statistical tasks such as hypothesis testing and assessment
of parameter uncertainty are possible, even though it becomes necessary
to adjust the number of degrees of freedom for the nonestimable
parameters. A complete description of this approach, which is rooted in
the theory of extended exponential families, is beyond the scope of the
article. See \citet{MLE} for details.

We can extend our study of the $\beta$-model in a number of ways. In
the supplementary material to this article, we consider various
generalizations of the $\beta$-model setting, including the
$\beta$-model with random numbers of edges, the Rasch model from item
response theory, the Bradley--Terry paired comparisons model and the
$p_1$ network model. For most of these models we were able to carry out
a fairly explicit analysis based on the underlying geometry, but for
the full $p_1$ model the complexity of the model polytope appears to
make such a direct analysis very difficult [this is reflected in
the high complexity of the Markov basis for $p_1$ model, of which we
give full account in \citet{p1markov}]. Another interesting
extension of
our results of Section~\ref{secasymptotics} would be to translate our
conditions, which are formulated in terms of expected degree sequences,
into conditions on the $p_{i,j}$'s themselves, for instance, by
establishing appropriate bounds for $\min_{i<j} p_{i,j}$, $\max_{i<j}
p_{i,j}$ or $\max_{i \neq j} \frac{p_{i,j}}{1 - p_{i,j}}$.

We conclude with some remarks on the computational aspects of our
analysis, which constitute a nontrivial component of our work and is of
key importance for detecting the nonexistence of the MLE and
identifying estimable parameters. The main difficulty in applying our
results is that the polytope of degree sequences $P_n$ is difficult to
handle algorithmically in general. Indeed, $P_n$ arises a Minkowksi sum
and, even though the system of defining inequalities is given
explicitly, its combinatorial complexity grows exponentially in $n$.
More importantly, the vertices of $P_n$ are not known explicitly.
Algorithms for obtaining the vertices of $P_n$, such as
\texttt{minksum} [see \citet{minksum}], are
computationally\vspace*{1pt}
expensive and require generating all the points $\{ \mathrm{A} x, x
\in\mathcal{G}_n\}$, where $|\mathcal{G}_n| = 2^{n \choose2}$, a task
that, even for $n$ as small as~$10$, is impractical. See, for instance,
our analysis of the $p_1$ model included in the supplementary material.
Thus, deciding whether a given degree sequence is a point in the
interior of $P_n$ and identifying the facial set corresponding to an
observed degree sequence on its boundary is highly nontrivial. Our
strategy to overcome these problems entails re-expressing the
$\beta$-model as a log-linear model with ${n \choose2}$
product-multinomial sampling constraints. This approach is not new, and
it harks back to the earlier re-expression of the Holland--Leinhardt
$p_1$ model and its natural generalizations as log-linear models
[\citet{fienmeyewass1985}, Fienberg and Wasserman
(\citeyear{fienwass1981,fienwass1981a}), \citet{meye1983}].
Though this re-parametrization increases the dimensionality of the
problem, it nonetheless has the crucial computational advantage of
reducing the determination of the facial sets of $P_n$ to the
determination of the facial sets of a pointed polyhedral cone spanned
by $n(n-1)$ vectors, which is a much simpler object to analyze, both
theoretically and algorithmically. This procedure is known as the
Cayley embedding in polyhedral geometry, and \citet{MLE} describe
its use in the analysis of log-linear models. The advantages of this
re-parametrization are two-fold. First, it allows us to use the highly
optimized algorithms available in \texttt{polymake}
[\citet{polymake}] for listing
explicitly all the facial sets of $P_n$. This is how we computed the
facial sets in all the examples presented in this article. Second, the
general algorithms for detecting nonexistence of the MLE and
identifying facial sets proposed in \citet{MLE}, which can handle
larger-dimensional models (with $n$ in the order of hundreds), can be
directly applied to this problem. This reference is also relevant for
dealing with inference under a nonexistent MLE.

The details of our computations and the associated algorithms are
provided in the supplementary material accompanying this article.
The \texttt{R} routines used to carry out the computations for the results
presented in the paper and for creating the input files for
\texttt{polymake} are available at
\href{http://www.stat.cmu.edu/\textasciitilde arinaldo/Rinaldo\_Petrovic\_Fienberg\_Rcode.txt}%
{http://}
\href{http://www.stat.cmu.edu/\textasciitilde arinaldo/Rinaldo\_Petrovic\_Fienberg\_Rcode.txt}%
{www.stat.cmu.edu/\textasciitilde arinaldo/Rinaldo\_Petrovic\_Fienberg\_Rcode.txt}.

\section{Proofs}

\mbox{}

\begin{pf*}{Proof of Theorem~\ref{thmmle}}
Throughout the proof, we will use standard results and terminology from
the theory of exponential families, for which standard references are
\citet{BROWN86} and \citet{BN78}.
The polytope
\[
S_n: = \operatorname{convhull} \bigl( \{ \mathrm{A} x, x \in
\mathcal{S}_n \} \bigr)
\]
is the convex support for the sufficient statistics of the natural
exponential family described in Section~\ref{secintro}. Furthermore,
by a fundamental result in the theory of exponential families
[see, e.g., Theorem 9.13 in \citet{BN78}], the MLE of the natural
parameter $\beta\in\mathbb{R}^n$ [or, equivalently of the set
probabilities $\{ p_{i,j}, i < j\} \in\mathbb{R}^{{n
\choose2}}$\vadjust{\goodbreak}
satisfying (\ref{eqpij})] exists if and only if $d \in\operatorname{int}(S_n)$.
Thus, it is sufficient to show that $d \in\operatorname{int}(S_n)$ if
and only if
$\tilde{d} \in\operatorname{int}(P_n)$.

Denote with $a_{i,j}$ the column of $\mathrm{A}$ corresponding to the
ordered pair $(i,j)$, with $i<j$, and set
%
%
\begin{equation}
\label{eqPij} P_{i,j} = \operatorname{convhull}\{ 0, a_{i,j}\}
\subset\mathbb{R}^n.
\end{equation}
Each $P_{i,j}$ is a line segment between its vertices $0$ and $a_{i,j}$.
Then, $P_n$ can be expressed as the zonotope obtained as the Minkowski
sum of the line segments $P_{i,j}$,
%
%
\begin{equation}
\label{eqPnsum} P_n = \sum_{i<j}
P_{i,j}.
\end{equation}
This identity can be established as follows. On one hand, $P_n$ is the
convex hull of vectors that are Boolean combinations of the columns of
$\mathrm{A}$. Since all such combinations are in $\sum_{i<j} P_{i,j}$, and
both $P_n$ and $\sum_{i<j} P_{i,j}$ are closed sets, we obtain $P_n
\subseteq\sum_{i<j} P_{i,j}$. On the other hand, the vertices of
$\sum_{i<j} P_{i,j}$ are also\vspace*{1pt} Boolean combinations of the columns of
$\mathrm{A}$ [see, e.g., Corollary 2.2 in \citet{FUKUDA04}], and,
therefore, $\sum_{i<j} P_{i,j} \subseteq P_n$.

Equation (\ref{eqPnsum}) shows, in particular, that $\tilde{d} \in
P_n$. Furthermore, using the same arguments, we see that, similarly to
$P_n$, $S_n$ too can be expressed as a Minkowski sum,
\[
S_n = \sum_{i<j}S_{i,j},
\]
where
\[
S_{i,j}: = P_{i,j} N_{i,j} = \{ x N_{i,j}
\dvtx x \in P_{i,j}\}
\]
is the rescaling of $P_{i,j}$ by a factor of $N_{i,j}$. In fact, we
will prove that $S_n$ and $P_n$ are combinatorially equivalent.

For a polytope $P$ and a vector $c$, we set $F(P;c): = \{ x \in P
\dvtx x^\top c \geq y^\top c, \forall y \in P\}$. Any face $F$ of $P$
can be written in this way, where $c$ is any vector in the interior of
the normal cone to $F$.
By Proposition 2.1 in \citet{FUKUDA04}, $F$ is a face of $P_n$
with $F
= F(P_n,c)$ if and only if it can be written uniquely as
\[
F(P_n,c) = \sum_{i < j}
F(P_{i,j},c)
\]
for any $c$ in the interior of the normal cone to $F$.
It is immediate to see that
$F(P_{i,j},c)$ is a face of $P_{i,j}$
if and only if $F(S_{i,j},c)$ is a face of $S_{i,j}$, and that
$F(S_{i,j},c) = N_{i,j} F(P_{i,j},c)$; in fact, $P_{i,j}$ and $S_{i,j}$
are combinatorially equivalent. Therefore, invoking again Proposition
2.1 in \citet{FUKUDA04}, we conclude that $F(P_{i,j},c)$ is a face of
$P_n$ if and only if
\[
\sum_{i < j} N_{i,j} F(P_{i,j},c)
\]
is a face of $S_n$ (and this representation is unique). From this, we
see that $P_n$ and $S_n$ have the same normal fan and, therefore, are
combinatorially equivalent.
\end{pf*}

\begin{pf*}{Proof of Lemma~\ref{corfacial}}
By Proposition 2.1 in \citet{FUKUDA04},
%
%
\begin{equation}
\label{equniq} F = F(P_n,c) = \sum_{i < j}
F(P_{i,j},c)
\end{equation}
for any $c$ in the interior of the normal cone to $F$, where the above
representation is unique. Since $P_{i,j}$ is a line segment [see (\ref
{eqPij})], its only proper faces are the vertices $0$ and $a_{i,j}$.
Let the set $\mathcal{F}$ be the complement of the set of pairs
$(i,j)$ with $i<j$ such that $F(P_{i,j},c)$ is either the vector $0$ or
$a_{i,j}$. By the uniqueness of the representation~(\ref{equniq}),
$\mathcal{F}$ is unique as well and, in particular, maximal.
Furthermore, as it depends on $F$ only through the interior of its
normal cone and since the interiors of the normal cones of $P_n$ are
disjoint, different faces will be associated with different facial sets.
\end{pf*}

\begin{pf*}{Proof of Theorem~\ref{thmus}}
Let $\tilde{d} = (\tilde{d}_1,\ldots,\tilde{d}_n)$ be the random
vector defined in (\ref{eqdtilde}). We will show that, under the
stated assumptions, $\tilde{d} \in\operatorname{int}(P_n)$ with
probability no
smaller than $1 - \frac{2}{n^{2c-1}}$.

Since $N$ is constant, we conveniently re-express the random vector
$\tilde{d}$ as an average of independent and identically distributed
graphical degree sequences. In detail, we can write
%
%
\begin{equation}
\label{eqdk} \tilde{d} = \frac{1}{N}\sum_{k=1}^N
d^{(k)},
\end{equation}
where each $d^{(k)}$ is the degree sequence arising from of an
independent realization of random graph with edge probabilities $\{
p_{i,j} \dvtx i < j \}$, for $k=1,\ldots,N$.

Thus, each $\tilde{d}_i$ is the sum of $N(n-1)$ independent random
variables taking values in $\{ 0,\frac{1}{N} \}$. Then, an application
of Hoeffding's inequality and of the union bound yields that the event
%
%
\begin{equation}
\label{eqhoef} \mathcal{O}_n: = \biggl\{ {\max_i}
|\tilde{d}_i - \overline{d}_i| \leq\sqrt{ c
\frac{n \log n}{N}} \biggr\}
\end{equation}
occurs with probability at least $1 - \frac{2}{n^{2c-1}}$.
Throughout the rest of the proof we assume that the event $\mathcal
{O}_n$ holds.

By assumption (i), for each $i$,
\begin{eqnarray*}
0 &<& C + \sqrt{ c \frac{n \log n}{N}} \leq\overline{d}_i - \sqrt{ c
\frac{n \log n}{N}} \leq\tilde{d}_i \leq\overline{d}_i +
\sqrt{ c \frac{n \log n}{N}} \\
&\leq& n-1 - C - \sqrt{ c \frac{n \log
n}{N}} < n-1,
\end{eqnarray*}
so that
%
%
\begin{equation}
\label{eqdi0} 0 < \tilde{d}_i < n-1,\qquad i=1,\ldots,n.
\end{equation}

Notice that the assumed constraint on the range of $C$ guarantees the
above inequalities are well defined.
Next, for each pair $(S,T) \in\mathcal{P}$,
\[
\bigl|g(S,T,\tilde{d},n) - g(S,T,\overline{d},n)\bigr| \leq{|S \cup T| \max
_i} |\tilde{d}_i - \overline{d}_i|,
\]
which yields
\[
g(S,T,\tilde{d},n) \geq g(S,T,\overline{d},n) - |S \cup T| \sqrt{ c
\frac{n \log n}{N}}.
\]
Using assumption (ii), the previous inequality implies that
%
%
\begin{equation}
\label{eqF} \min_{(S,T) \in\mathcal{P}} g(S,T,\tilde{d},n) > C > 0.
\end{equation}
Thus, we have shown that (\ref{eqdi0}) and (\ref{eqF}) hold,
provided that the event $\mathcal{O}_n$ is true and assuming (i)
and (ii). Therefore, by Theorem~\ref{thmMP} the MLE exists.
\end{pf*}

\begin{pf*}{Proof of Corollary~\ref{corus}}
Using the same setting and notation of Theorem~\ref{thmus}, we will
assume throughout the proof that the event
\[
\mathcal{O}'_n:= \Bigl\{ \max_k
\max_i \bigl|d^{(k)}_i - \overline
{d}_i\bigr| \leq\sqrt{c n \log n} \Bigr\}
\]
holds true. By Hoeffding's inequality, the union bound and the
inequality $\log N \leq\log n$, we have
\[
\mathbb{P}\bigl(\mathcal{O}'^c_n\bigr)
\leq2 \exp\{ - 2c \log n + \log n + \log N \} \leq\frac{2}{n^{2c - 2}}.
\]
A simple calculation shows that, when $\mathcal{O}'_n$ is satisfied,
we also have
\[
\Bigl\{ {\max_i }|\tilde{d}_i -
\overline{d}_i| \leq\sqrt{ c n \log n} \Bigr\}.
\]
Then, by the same arguments we used in the proof of Theorem \ref
{thmus}, assumption (i$'$) yields that
%
%
\begin{equation}
\label{eqdi0v2} 0 < \tilde{d}_i < n-1,\qquad i=1,\ldots,n,
\end{equation}
and, for each pair $(S,T) \in\mathcal{P}$,
%
%
\begin{equation}
\label{eqohmanv2} g(S,T,\tilde{d},n) \geq g(S,T,\overline{d},n) - |S
\cup T|
\sqrt{ c n \log n}.
\end{equation}

It is easy to see that, for the event $\mathcal{O}'_n$, assumption
(i$'$) also yields
%
%
\begin{equation}
\label{eqdiv2} \min_k \min_{i} \min
\bigl\{ d^{(k)}_i, n-1-d^{(k)}_i \bigr
\} \geq\sqrt{ c n \log n} + C.
\end{equation}

We now show that, when (\ref{eqdi0v2}) and the previous equation are
satisfied, the MLE exists if
%
%
\begin{equation}
\label{eqFnv2} \min_{(S,T) \in\mathcal{P}_n} g(S,T,d,n) > C > 0.
\end{equation}
Indeed, suppose that (\ref{eqdi0v2}) is true and that $\tilde{d}$
belongs to the boundary of~$P_n$.
Then, by the integrality of the polytope $P_n$, there exist nonempty
and disjoint subsets $T$ and $S$ of $\{1,\ldots,n\}$ satisfying the
conditions of Lemma~\ref{lemMP} for each of the degree sequences
$d^{(1)},\ldots,d^{(k)}$. If $\min_k \min_i d^{(k)}_i > \sqrt{c n
\log n} + C$, then, necessarily, $|S| > \sqrt{c n \log n} + C$,
because $|S|$ is the maximal degree of every node $i \in T$. Similarly,
since each $i \in S$ has degree at least $|S| - 1 + |(S \cup T)^c|$, if
$\max_k \max_i d^{(k)}_i < n - 1 -\sqrt{c n \log n} - C$, the inequality
\[
|S| - 1 + \bigl|(S \cup T)^c\bigr| < n -1 - \sqrt{c n \log n} - C
\]
must hold, implying that $|T| = n - |S| - |(S \cup T)^c| > \sqrt{c n
\log n} + C$. Thus, we have shown that if (\ref{eqdi0v2}) and (\ref
{eqdiv2}) hold, and $\tilde{d}$ belongs to the boundary of~$P_n$, the
cardinalities of the sets $S$ and $T$ defining the facet of $P_n$ to
which $\tilde{d}$ belongs cannot be smaller than $\sqrt{c n \log n} +
C$. By Theorem~\ref{thmMP}, when (\ref{eqdi0v2}) and (\ref
{eqdiv2}) hold, (\ref{eqFnv2}) implies that $\tilde{d} \in
\operatorname
{int}(P_n)$, so the MLE exists. However, equation (\ref{eqohmanv2})
and assumption (ii$'$) implies (\ref{eqFnv2}), so the proof is
complete.~%
\end{pf*}

\section*{Acknowledgments}

A previous version of this manuscript was completed while the second
author was in residence at Institut Mittag-Leffler,
for whose hospitality she is grateful.

\begin{supplement}
\stitle{Supplement to ``Maximum lilkelihood estimation in the $\beta$-model''}
\slink[doi]{10.1214/12-AOS1078SUPP} 
\sdatatype{.pdf}
\sfilename{aos1078\_supp.pdf}
\sdescription{In the supplementary material we extend our analysis to
other models for network data: the Rasch model, the $\beta$-model with
no sampling constraints on the number of observed edges per dyad, the
Bradley--Terry model and the $p_1$ model of \citet{HL81}.
We also provide details on how to determine whether a given degree
sequence belongs to the interior of the polytope of degree sequences
$P_n$ and on how to compute the facial set corresponding to a degree
sequence on the boundary of $P_n$.}
\end{supplement}

%

\printaddresses

\end{document}